\documentclass[epj]{svjour}
\usepackage{graphics}
\usepackage{graphicx}
\usepackage{amsmath}
\usepackage{amssymb}
\usepackage{amssymb,bbold}
\RequirePackage[colorlinks,citecolor=blue,urlcolor=blue,linkcolor=blue]{hyperref}
\usepackage{subfig}
\journalname{EPJ A}

\begin{document}
\title{Non extensive thermodynamics and neutron star properties}
\author{D\'ebora P. Menezes\inst{1} \and Airton Deppman\inst{2} \and Eugenio Meg\'{\i}as\inst{3,4} \and
Luis B. Castro\inst{5}
}                     
%
%
\institute{Departamento de F\'{\i}sica - CFM - Universidade Federal de Santa Catarina,
Florian\'{o}polis - SC - CP. 476 - CEP 88.040 - 900 - Brazil, \email{debora.p.m@ufsc.br} \and 
Instituto de F\'isica, Universidade de S\~ao Paulo - Rua do
Mat\~ao Travessa R Nr.187 CEP 05508-090 Cidade Universit\'aria, S\~ao Paulo -
Brazil, \email{deppman@if.usp.br} \and
Grup de F\'{\i}sica Te\`orica and IFAE, Departament de
F\'{\i}sica, Universitat Aut\`onoma de Barcelona, Bellaterra E-08193
Barcelona, Spain,
\and
Max-Planck-Institut f\"ur Physik (Werner-Heisenberg-Institut), F\"ohringer Ring 6, D-80805, Munich, Germany, \email{emegias@mppmu.mpg.de} \and
Departamento de F\'{\i}sica, Universidade Federal do Maranh\~{a}o, Campus Universit\'{a}rio do Bacanga, CEP 65080-805, S\~{a}o Lu\'{i}s, MA, Brazil, \email{lrb.castro@ufma.br}
}
\date{Received: date / Revised version: date}
%
\abstract{
In the present work we apply non extensive statistics to obtain
equations of state suitable to describe stellar matter and verify its
effects on microscopic and macroscopic quantities. Two snapshots of
the star evolution are considered and the direct Urca process is
investigated with  two different parameter sets. $q$-values are
chosen as 1.05 and 1.14.
The equations of state are only slightly modified, but
the effects are enough to produce stars with slightly higher maximum
masses. The onsets of the constituents are more strongly affected and
the internal stellar temperature decreases with the increase of the
$q$-value, with consequences on the strangeness and cooling rates of the stars.
} 
\maketitle
\section{Introduction}
\label{intro}
A type II supernova explosion is triggered when
massive stars ($8~M_\odot < M < 30~M_\odot$ ) exhaust their fuel
supply, causing the core to be crushed by gravity.
The remnant of this gravitational collapse is a compact star or a black hole, depending on the
initial condition of the collapse. Newly-born protoneutron stars (PNS) are hot
and rich in leptons, mostly $e^-$ and $\nu_e$  and have masses of the order of
$1-2~M_\odot$ \cite{Lattimer2004,keil1995}. During the very beginning of the
evolution, most of the binding energy, of the order of $10^{53}$ ergs is
radiated away by the neutrinos.  During the temporal evolution of the
PNS in the so-called Kelvin-Helmholtz epoch, the remnant compact
object changes from a hot and lepton-rich PNS to a cold and
deleptonized neutron star \cite{pons}.
The neutrinos already present or generated in the PNS hot matter escape by
diffusion  because of the very high densities and temperatures involved.
At zero temperature no trapped neutrinos are left in the star because their
mean free path would be larger than the compact star radius.
Simulations have shown that the evolutionary picture can be understood if one studies three snapshots of the time evolution of a compact star in
its first minutes of life \cite{Prakash:1996xs}.  At first, the PNS is
warm (represented by fixed entropy per particle) and has a large number of
trapped neutrinos (represented by fixed lepton fraction). As the trapped
neutrinos diffuse, they heat up the star. Finally, the star
is considered cold.

To describe these three snapshots, appropriate equations of state
(EOS) have to be used. These EOS are normally parameter dependent and
are adjusted so as to reproduce nuclear matter bulk properties, as the
binding energy at the correct saturation density and incompressibility
as well as ground state properties of some nuclei \cite{sw,glen00,Haensel}.
Until recently, when two stars with masses of the order of 2$M_\odot$ were
confirmed \cite{demorest,antoniadis}, most EOS  were expected to produce
maximum stellar  masses just larger than 1.44$M_\odot$ and radii of the order
of 10 to 13 km. The new measurements imposed more rigid constraints on
the EOS.

On the other hand, the effects of non extensive statistical
mechanics \cite{Tsallis88,TsallisBook} have been explored both in
high-energy physics \cite{Bediaga,Deppman12} and astrophysical
problems \cite{lavagno}. The $q$-deformed entropy functional that
underlines non extensive statistics depends on a real parameter ($q$)
that determines the degree of nonadditivity of the functional and in
the limit $q \rightarrow 1$, it becomes additive and the standard
Boltzmann-Gibbs entropy is recovered. The results of High Energy Physics (HEP) experiments have shown that non extensive statistics can play an important role in the description of collisions with energy above
$\sqrt{s} \sim$~10~GeV~\cite{Bediaga,Beck,CW}. In fact the well-known Hagedorn's theory~\cite{Hagedorn} can be extended to include non extesive statistics, resulting in a Non Extensive Self-Consistent Thermodynamics (NESCT) predicting a limiting temperature, $T_o$, a characteristic entropic index, $q_o$, for the hot hadronic system obtained at HEP experiments and a new hadron mass spectrum formula.

Systematic analyses of HEP data have shown that indeed a limiting temperature is obtained, with $T_o= 61$~GeV and $q_o=1.14$~\cite{CW,Lucas,Lucas2,Sena}. In addition it was shown that the new hadron mass spectrum formula describes very well the known hadronic states masses with values for $T_o$ and $q_o$ that are in agreement with those found in HEP data analysis. These results show that the main aspects of high energy collisions can be described by the NESCT approach.

With the values for $T_o$ and for $q_o$ one obtains the entire thermodynamical description according to the non extensive theory for null chemical potential,~$\mu$, as expected to happen in HEP. In fact most of the Lattice QCD (LQCD) calculations are perfomed with $\mu=0$. A comparison between NESCT and LQCD results was done in~\cite{Deppman2} showing that there was a fair agreement between the results from both approaches considering the large differences in the results from different LQCD calculations. Notice that LQCD does not include non extensivity explicitly, so a conclusion one can get from here is that non extensivity is an emerging feature for QCD interacting systems.

The extension of NESCT to finite chemical potential was performed in~\cite{ours} (see also~\cite{Megias:2014tha,Deppman:2015cda}), where it was also obtained the partition function for a non extensive quantum ideal gas. This work opens the possibility to use NESCT in systems very different from those in HEP. The study of neutron stars, where low temperature hadronic matter at extremely high densities can give rise to a phase transition that is in many aspects similar to that observed in HEP experiment, is a potential candidate.

In the present work we investigate how the consideration of
non extensive statistics affects hadronic matter at finite temperature
and large densities by applying it to PNS. Based on an extensive study
of parameter dependent relativistic models~\cite{jirina} and on the
mentioned 2$M_\odot$ stars, we have opted to work with two
parametrizations of the non-linear Walecka model~\cite{sw}, namely GM1~\cite{gm1} and IUFSU~\cite{iufsu}. Hence, we also check how parameter
dependent the stellar matter microscopic (EOS, particle fractions,
strangeness, internal temperature, direct Urca process onset) and
macroscopic (radius, gravitational and baryonic masses, central energy
density) properties are when Tsallis statistics is used.

The work is organized as follows: in Section~\ref{formalism}, the
basic equations necessary to follow the EOS calculations both with
standard hadrodynamics and with non extensive statistics are outlined.
In Section~\ref{results} our results are displayed and discussed, and finally the main conclusions are drawn in Section~\ref{conclusion}.

\section{The Formalism}
\label{formalism}

\subsection{Standard quantum hadrodynamics}
\label{subsec:quantum_hadrodynamics}

In this section we present the hadronic equations of state (EOS)
used in this work. We describe hadronic matter
within the framework of the relativistic non-linear Walecka model (NLWM)
\cite{sw}. In this  model the nucleons are coupled to
neutral scalar $\sigma$,
isoscalar-vector $\omega_\mu$ and isovector-vector $\vec \rho_\mu$  meson fields.
We also include a $\rho-\omega$ meson coupling term as in
\cite{iufsu,hor01,fsu} because it was shown to have important
consequences in neutron star properties related to
the symmetry energy and its slope \cite{cp2014}.

The Lagrangian density reads
\begin{eqnarray}
{\cal L}&=& \sum_j {\bar \psi_j \left[ ~ { \gamma_\mu  \left( {i \partial^\mu
- g_{\omega j} \, \omega^\mu - g_{\rho j} \, \vec \tau _j \,.\, \vec \rho^{\, \mu}  } \right)
- m_j^* ~} \right]\psi_j }  \nonumber \\
&+& \frac{1}{2} {\partial_\mu \sigma \partial^\mu \sigma - \frac{1}{2} m_{\sigma}^2 \sigma^2 }
- \frac{1}{{3!}} k \sigma^3  - \frac{1}{{4!}} \lambda \sigma^4     \nonumber \\
&-& \frac{1}{4} \Omega_{\mu \nu} \, \Omega^{\mu \nu}
+ \frac{1}{2} m_{\omega}^2 \, \omega_\mu \omega^\mu
+ \frac{1}{4!}\xi g_{\omega}^4 (\omega_{\mu} \omega^{\mu})^2  \nonumber \\
&-& \frac{1}{4} \vec R_{\mu \nu } \,.\,  \vec R^{\mu \nu } +
\frac{1}{2} m_\rho^2 \, \vec \rho_{\mu} \,.\, \vec \rho^{\, \mu}   \nonumber \\
&+& \Lambda_{\rm v} (g_{\rho}^2 \; \vec \rho_{\mu} \,.\, \vec \rho^{\,
  \mu} )(g_{\omega}^2 \; \omega_{\mu} \omega^{\mu} )  \;   \nonumber \\
&+& \sum\limits_l {\bar \psi _l \left( {i\gamma _\mu  \partial ^\mu
- m_l } \right)\psi _l}  \;,
\label{baryon-lag}
\end{eqnarray}
\noindent where
\begin{equation}
m_j^* = m_j - g_{\sigma j} \sigma
\end{equation}
\noindent is the baryon effective mass,
$\Omega_{\mu\nu }=\partial_\mu \omega_\nu - \partial_\nu \omega_\mu$~,
$\vec R_{\mu \nu } = \partial_\mu \vec \rho_\nu -\partial_\nu \vec \rho_\mu
-g_\rho \left({\vec \rho_\mu \,\times \, \vec \rho_\nu } \right)$,
$g_{ij}=X_i g_i$
are the coupling constants of mesons $i = \sigma, \omega, \rho$ with baryon
$j$, $m_i, i=\sigma,\omega,\rho$ is the mass of meson $i$ and $l$ represents the leptons
$e^-$ and $\mu^-$ and respective neutrinos.
The couplings
$k$~($k = 2\,M_N\,g_{\sigma}^3\,b$) and $\lambda$~($\lambda = 6\, g_{\sigma}^4\,c$)
are the weights of the non-linear scalar terms, $\Lambda$ is the
weight of the cross $\omega-\rho$ interaction and $\vec \tau$ is the isospin
operator.
The sum over $j$ in (\ref{baryon-lag}) can be extended over neutrons
and protons only or over the  lightest eight  baryons
$\{ n,p,\Lambda,\Sigma^-,\Sigma^0,\Sigma^+,\Xi^-,\Xi^0 \}$.
The coupling constants
$\{ g_{\sigma j} \}_{j=\Lambda,\Sigma,\Xi}$ of the hyperons with the scalar
meson $\sigma$ can be constrained
by the hyper-nuclear potentials in nuclear matter to be consistent with hyper-nuclear
data \cite{hyp1,lopes2014}, but we next  consider $X_\sigma$=0.7 and
$X_\omega=X_\rho$=0.783 and equal for all the hyperons as in
\cite{glen00}. As it is well known that the softness/stiffness of
  the EOS dependes on the value of these unknown quantitites, we
  restrict ourselves just to one possible case.
In Table \ref{tab-parameters} we give the symmetric nuclear matter properties
at saturation density as well as the parameters of the models used in the
present work.

Applying the Euler-Lagrange equations to (\ref{baryon-lag}),
assuming translacional and rotational invariance, static mesonic fields
and using the mean-field approximation
($\sigma \to \langle \sigma \rangle = \sigma_{0}~ ;\;
\omega_\mu \to \langle \omega_\mu  \rangle = \delta_{\mu 0} \,\omega_0 ~;\;
\vec \rho_\mu  \to \langle \vec \rho_\mu  \rangle = \delta_{\mu 0} \, \delta^{i 3} \rho_0^3
\equiv \delta_{\mu 0} \, \delta^{i 3} \rho_{03}  $), we obtain the
following equations of motion for the meson fields:
\begin{equation}
m_{\sigma}^2 \, \sigma_{0} = - \frac{k}{2} \sigma _{0}^2 - \frac{\lambda }{6} \sigma _{0}^3
+ \sum\limits_j^{} g_{\sigma j} \, n_{j}^{s} \;, \nonumber
\label{sigma-field-boson1}
\end{equation}
\begin{align}
m_{\omega}^2 \, \omega_{0} &= -\dfrac{\xi g_{\omega}^4}{6} \omega_0^3
+ \sum\limits_j^{} g_{\omega j} \, n_{j}^{}
- 2 \Lambda_{\rm v} \, g_{\rho}^2 \, g_{\omega}^2 \, \rho_{03}^2 \,\, \omega_0 \;,
\nonumber \\
m_\rho ^2 \, \rho_{03} &= \sum\limits_j^{}  g_{\rho j}\, \tau_{3j}\, n_{j}^{}
- 2 \Lambda_{\rm v} \, g_{\rho}^2 \, g_{\omega}^2 \, \omega_0^2 \,\, \rho_{03} \;,
\label{eq-mov-rho03}
\end{align}
where
\begin{equation}
n_{j}^s = \int \frac{d^3 p}{(2 \pi)^3} \frac{{m_j^*}}{E_j^*}  (f_{j+}
+ f_{j-})  \;,
\label{scalarrho}
\end{equation}
\noindent is the baryon scalar density of particle $j$ and the respective baryon density
\begin{equation}
 n_{j} =  \frac{2}{(2 \pi)^3} \int\limits_{}^{}{d^3 p \, (f_{j+} -
   f_{j-}) }, \qquad n_B = \sum_j n_j ,
\label{densB}
\end{equation}
\noindent and $f_{j\pm}$ is the Fermi distribution for the baryons (+) and anti-baryons (-) $j$:
\begin{equation}
f_{j\pm} = \frac{1}{e^{\beta(E_j^* \mp \nu_j)} + 1},
\end{equation}
\noindent with $\beta = 1/T$, $E_j^* = ({\mathbf p}_j^2 + m_j^{*\,2})^{1/2}$
and the effective chemical potential of baryon $j$ is given by
\begin{equation}
  \nu_j = \mu_j - g_{\omega j}\omega_0 - \tau_{3j}  \, g_{\rho j}\, \rho_{03} \;.
\label{pot-quim-nucleons}
\end{equation}

The EOS can then be calculated and reads:
\begin{align}
P &= \frac{1}{3 \pi^2} {\sum\limits_{j}^{} {\int_{}^{ }}
{\frac{p^4 dp}{\sqrt {p^2 + m_j^{*2}}}}} (f_{j+} \,+\, f_{j-})
+ \frac{{m_\omega^2 }}{2} \omega_0^2+ \frac{\xi}{24} \omega_0^4   \nonumber \\
&+ \frac{{m_\rho ^2 }}{2} \rho_{03}^2
- \frac{m_\sigma^2}{2} \sigma_{0}^2 - \frac{k}{6} \sigma _{0}^3
- \frac{\lambda }{24} \sigma _{0}^4+ \Lambda_{\rm v} \, g_{\rho}^2 \, g_{\omega}^2 \, \omega_0^2 \, \rho_{03}^2  \nonumber \\
&+\frac{1}{3 \pi^2} {\sum\limits_l {\int_{}^{ }}
{\frac{p^4 dp}{\sqrt {p^2 + m_l^2}}}} (f_{l+} \,+\, f_{l-}) ,
\label{press-bar}
\end{align}

\begin{align}
{\cal E} &= \frac{1}{\pi^2}  {\sum\limits_{j}^{} {\int_{}^{} {p^2 dp \sqrt {p^2 + m_j^{*2}}}} (f_{j+} \,+\, f_{j-}) }
+ \frac{{m_\omega^2 }}{2} \omega_0^2+ \frac{\xi}{8} \omega_0^4 \nonumber \\
& + \frac{{m_\rho ^2 }}{2} \rho_{03}^2
+ \frac{{m_\sigma^2 }}{2} \sigma _{0}^2 + \frac{k}{6} \sigma _{0}^3
+ \frac{\lambda }{{24}} \sigma _{0}^4
+ 3 \Lambda_{\rm v} \, g_{\rho}^2 \, g_{\omega}^2 \, \omega_0^2 \, \rho_{03}^2 \qquad \nonumber \\
&+\frac{1}{\pi^2} {\sum\limits_l {\int_{}^{ }}
{p^2 dp \sqrt {p^2 + m_l^2}}} (f_{l+} \,+\, f_{l-}) ,
\label{deng-bar}
\end{align}
\noindent where the last terms in Eqs.(\ref{press-bar}) and
  (\ref{deng-bar}) are due to the inclusion of leptons as a free gas
  in the system and their
 distribution functions are given by
\begin{equation}
f_{l\pm} = \frac{1}{e^{\beta(E_l \mp \mu_l)} + 1},
\end{equation}
with $E_l= ({\mathbf p}_l^2 + m_l^2)^{1/2}$.

The entropy per particle (baryon) can be calculated through the thermodynamical expression
\begin{eqnarray}
\frac{\cal S}{n_B}=\frac{{\cal E}+P- \sum_j \mu_j n_j}{T n_B}.
\end{eqnarray}

When the hyperons are present we define the strangeness fraction:
\begin{equation}
 f_s = \frac{1}{3} \frac{\sum_{j}^{} |s_j| n_j }{n_B} \; ,
\end{equation}
where $s_j$ is the strangeness of baryon $j$ and $n_B$ is
the total baryonic density given in eq.~(\ref{densB}).

\begin{table*}[ht]
\centering
\begin{tabular}{cccc}
\hline
 &  {\bf IU-FSU} \cite{iufsu} &  {\bf GM1} \cite{gm1} \\
\hline
$n_0$ (fm$^{-3}$)          &   0.155     &    0.153 \\
$K$ (MeV)                  &    231.2    &   300 \\
$m^*/m$                    &    0.62      &   0.70  \\
$m$~(MeV)                  &    939     &   938 \\
-$B/A$ (MeV)               &    16.4    &   16.3 \\
${\cal E}_{\rm sym}$ (MeV) &    31.3    &   32.5  \\
$L$ (MeV)                  &    47.2    &   94 \\
\hline
$m_{\sigma}$ (MeV)         &    491.5   &   512 \\
$m_{\omega}$ (MeV)         &    782.5   &   783\\
$m_{\rho}$ (MeV)           &    763     &   770 \\
$g_{\sigma}$               &    9.971     &  8.910  \\
$g_{\omega}$               &    13.032     &  10.610 \\
$g_{\rho}$                 &    13.590  &  8.196 \\
$b$                        & \;\; 0.001800 \;\;   & \;\; 0.002947 \;\; \\
$c$                        &  0.000049   & -0.001070 \\
$\xi$                      &    0.03    &      0    \\
$\Lambda_{\rm v}$    & 0.046    &     0    \\
\hline
\end{tabular}
\caption{Parameter sets used in this work and corresponding saturation
properties.}
\label{tab-parameters}
\end{table*}

\subsection{Non extensive statistics}
\label{subsec:nonextensive}

In order to introduce non extensivity in the NS problem we use the NESCT approach in obtaining the EOS for the hadronic matter. The extension for finite chemical potential given in Ref.~\cite{ours} is the most appropriate framework since we expect $\mu \neq 0$ for the NS matter. The starting point is the partition function~\cite{ours}
\begin{equation}
\begin{split}
\log\Xi_q(V,T,\mu) & =
 -\xi V\int \frac{d^3p}{{(2\pi)^3}} \sum_{r=\pm}\Theta(r x) \\
 & \times\log^{(-r)}_q\bigg(\frac{ e_q^{(r)}(x)-\xi}{ e_q^{(r)}(x)}\bigg) \,, \label{partitionfunc}
\end{split}
\end{equation}
\noindent where $x= \beta( E_p - \mu)$, we take $\xi = \pm 1$ for bosons and fermions respectively, $\Theta$ is the step function, and the q-logarithm
\begin{equation}
\begin{cases}
& \log^{(+)}_q(x)=\frac{x^{q-1}-1}{q-1}, \quad  x \geq 0,  \\
&  \log^{(-)}_q(x)=\frac{x^{1-q}-1}{1-q},  \quad x<0\,
\end{cases}
\end{equation}
\noindent is the inverse function of the q-exponential given by
\begin{equation}
\begin{cases}
 & e_q^{(+)}(x)=[1+(q-1)x]^{1/(q-1)} \qquad\;\;\,\,\,\,\,\,\,\,\,\,\,\,\,\,\,\,\,\,\,\,\,\,\,\,\,, \; x\geq 0\,, \\
 & e_q^{(-)}(x)=\frac{1}{ e_q^{(+)}(|x|)}=[1+(1-q)x]^{1/(1-q)}\qquad\,, \; x<0\,. \label{qexp}
\end{cases}
\end{equation}
From the definition of $q$-deformed entropy \cite{Plastino}, we can
write the distribution functions:
\begin{equation}
\begin{cases}
& n_q^{(+)}(x)=\frac{1}{(e_q^{(+)}(x) +1)^q}, \quad  x \geq 0,\\
& n_q^{(-)}(x)=\frac{1}{(e_q^{(-)}(x) +1)^{2-q}},\quad x<0. \label{eq:n}
\end{cases}
\end{equation}
From here one gets the entropy density
\begin{eqnarray}
 {\cal S} &=& \frac{1}{\pi^2(q-1)} \sum_j \sum_{r=\pm} \int p^2 dp  \, \Theta(r x_j) r \nonumber \\
&&\qquad \times \bigg[ 1- n_q^{(r)}(x_j) - \left(1 - \tilde{n}_q^{(r)}(x_j)\right)^{\tilde{q}} \bigg]\,,  \label{entropy}
\end{eqnarray}
where
\begin{equation}
\tilde{q}=
\begin{cases}
& q \qquad\quad\;\;\, \,,\,\,\,x\geq 0\,, \\
& 2-q \qquad \,,\,\,\,x<0\,,
\end{cases}
\end{equation}
and we have defined $\tilde{n}_q^{(\pm)}(x) \equiv 1/(e_q^{(\pm)}(x) +1)$. 

Before analysing the non extensive thermodynamics applied to a
  stellar system  it is worthwhile to discuss some differences between 
the approach used here and the one used in Ref.~\cite{lavagno}. The
$q$-exponential  functions defined in Eq.~(\ref{qexp}) are the 
same as the ones used by Lavagno and Pigato~\cite{lavagno}, but the
distribution function derived from the partition function 
adopted in the present work, as shown in Eq.~(\ref{eq:n}), differs
from the corresponding function in Ref.~\cite{lavagno} in the region
$x<0$.  Here the exponent in the denominator is $2-q$ while in their
work Lavagno and Pigato used the exponent $q$. 
There are in addition some typos in~\cite{Plastino}, as discussed
in~\cite{ours},  which remain unmodified in~\cite{lavagno}.
 It is important to notice that Eq.~(\ref{eq:n}) is consistently
 obtained from the partition function and entropy proposed in
 \cite{ours}. These comments refer to the regime $q>1$. The case $q<1$ is not discussed in details in the present work, but we give some insight at the end of this section. An interesting analysis of the several non extensive
 versions of a quantum ideal gas has been recently done in 
Ref.~\cite{rozynek}.

Notice that the distribution function $n_q^{(-)}(x)$ is a direct result of the application of the usual formalism of Thermodynamics to the proposed partition function, and the exponent $2-q$ is not introduced deliberately, but results from the usual calculations. Therefore $q$ in the equation for the distribution function has the same value as in other parts of the paper. It is worth to mention that there are recent approaches to this problem that avoids the discontinuity in the second derivatives of thermodynamical functions~\cite{Biro2015}. 

The pressure is
\begin{equation}
P = \frac{T}{\pi^2}  \sum_j \sum_{r=\pm} \int p^2 dp \, \Theta(r x_j) \log^{(-r)}_q\bigg( \frac{1}{1-\tilde{n}^{(r)}_q(x_j)} \bigg) \,, \label{press-tsallis}
\end{equation}
the baryonic density
\begin{equation}
{\cal N}=
\begin{cases}
& \frac{1}{\pi^2} \sum_j
\int p^2 dp  \, n_q^{(+)}(x_j)\,, \quad x_j \geq 0, \\
& \frac{1}{\pi^2} \sum_j
\int p^2 dp \, n_q^{(-)}(x_j) + 2 C_n\,, \quad x_j < 0
\end{cases}
\label{dens-tsallis}
\end{equation}
with
$$C_n= \frac{\mu_j T \sqrt{\mu_j^2 - {M_j^*}^2}}{2 \pi^2}
\frac{(2^{q-1}+2^{1-q}-2)}{q-1} \theta( \mu_j - M^*_j) $$
and the energy density:
\begin{equation}
{\cal E}=
\begin{cases}
& \frac{1}{\pi^2} \sum_j
\int p^2 dp ~E~~ n_q^{(+)}(x_j)\,, \quad x_j \geq 0, \\
& \frac{1}{\pi^2} \sum_j
\int p^2 dp  ~E~~ n_q^{(-)}(x_j) + 2 C_e\,, \quad x_j < 0,
\end{cases}
\label{ener-tsallis}
\end{equation}
with
$$
C_e = \mu_j C_n
$$
and where $x_j=\beta(E_j^* - \mu_j)$.

The constants $C_n$ and $C_e$ were introduced in Ref~\cite{ours}
  to tackle the jump in $n^{(\pm)}(x)$ at $x=0$.
 As observed in Ref.~\cite{rozynek} such a jump could be related to
 the excess of particles  and the deficiency of anti-particles at the border of the Fermi
surface, what is not observed at high energy. 
For a deeper discussion on this regard the reader is addressed to
Ref.~\cite{rozynek}. It is also worth noting that in the numerical
results presented next, these constants play practically no role.

When non extensive statistical mechanics is used instead of the usual
Fermi-Dirac expressions for the gas part of the EOS presented in
  the last section, the expressions
for pressure and energy density are rewritten in such
a way that the first and last terms in equations (\ref{press-bar}) and
(\ref{deng-bar}) are substituted by equations (\ref{press-tsallis})
and (\ref{ener-tsallis}) respectively. Moreover, the usual baryonic
density given in eq.~(\ref{densB}) is replaced by
eq.~(\ref{dens-tsallis}). In the equations of motion, the scalar
density eq.~(\ref{scalarrho}) is replaced by
\begin{equation}
{n^s_j}=
\begin{cases}
& \frac{1}{\pi^2} 
\int p^2 dp  \frac{m^*_j}{E^*_j}
n_q^{(+)}(x_j) \qquad x_j \geq 0, \\
& \frac{1}{\pi^2} 
\int p^2 dp  \frac{m^*_j}{E^*_j}  n_q^{(-)}(x_j) . \qquad x_j < 0
\end{cases}
\label{scalar-tsallis}
\end{equation}

Notice that thermodynamical consistency, shown in
  Ref.~\cite{ours} for non-interaction particles, is also achieved in the presence of interaction hadrons.

\subsubsection{Super and sub-extensive regimes}

Before we proceed by applying the above results to neutron stars,
 it is important to comment on possible choices for the $q$ value.
We can expand the entropy as
${\cal S}(q) = {\cal S}(q=1) + O(q-1) + O( (q-1)^2 ) + \cdots + O((q-1)^4) + \cdots $,
which is possible to compute for both $q$ larger and smaller than 1.

\begin{figure}[tbp]
\centering
\subfloat[]{\includegraphics[width=0.68\linewidth]{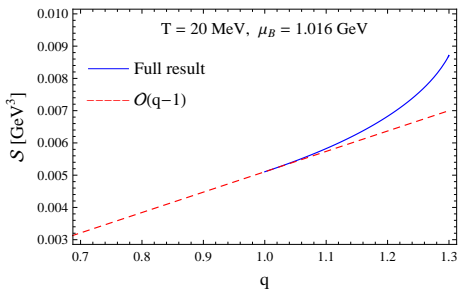}\label{fig0:a}}
\hfill
\subfloat[]{\includegraphics[width=0.68\linewidth]{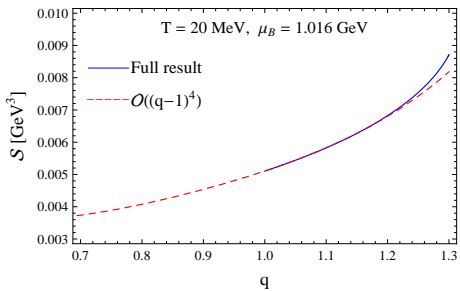}\label{fig0:b}}
\caption{Entropy density for different $q$ values with a fixed
temperature of $T=20$ MeV and a fixed chemical potential $\mu_B=1.016$ GeV.
The continuous blue line represents the full results and the dashed red lines the expansions
  up to a) $(q-1)$ and b)  $(q-1)^4$.
\label{fig0} }
\end{figure}

In order to exemplify the results, we add two figures for a
fixed temperature of $T=20$~MeV and fixed chemical potential
$\mu_B=1.016$ GeV. This temperature is chosen because it is of
  interest in the applications to protoneutrostars that follows. It
  was obtained in~\cite{ours} that at this value of the chemical
  potential a chemical freeze-out takes place at $T=20$~MeV. In
Fig.~\ref{fig0:a}, we compare the results for the entropy density with
$q>1$ and full computation, obtained from Eq.~(\ref{entropy}), with
the results obtained from the expansion above for $q>1$ and $q<1$ up
to order $(q-1)$. In Fig. \ref{fig0:b} the expansion goes up to order
$(q-1)^4$, showing a very good agreement with the full result.
For $q$ values lower than 1, the full computation would give complex
results, but the expansion would still be possible. Had we
decided to use $q<1$, as in \cite{lavagno}, the thermodynamic
quantities would have to be expanded, at least, up to order $q-1$.

\begin{figure}[tbp]
\centering
\subfloat[]{\includegraphics[width=0.68\linewidth]{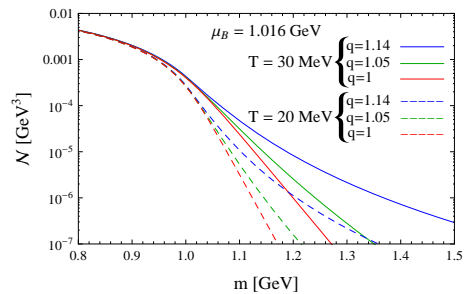}}
\caption{Baryonic density as a function of baryon mass for different $q$ values, $T=20$ and 30~MeV and $\mu_B=1.016$ GeV.
\label{fig00} }
\end{figure}

We also display in Fig.~\ref{fig00} the total baryonic
density ${\cal N}$, defined in Eq.~(\ref{dens-tsallis}), for different values of $q$ as a function
of the particle mass for fixed temperatures and the same chemical
potential as in the graphs for the entropy. Up to a certain mass, of
the order of 1.05 GeV at $T=20$ MeV and 1.1 GeV at $T=30$ MeV,
non extensivity plays almost no role, but as $q$ and $T$ increase,
heavier particles are favored. Of course, the value of the chosen
  chemical potential defines the mass value for which non extensivity
  becomes important and the chosen chemical potential is of the order
  of the baryon chemical potencials of the particles in stellar medium
  that will be investigated in the next sections.

\subsection{Stellar matter}
\label{stellarmatter}

In stellar matter there are two conditions that have to be
fulfilled, namely,
charge neutrality and $\beta$-stability and they read:
\begin{equation}
\sum_j q_j n_j + \sum_l q_l n_l =0,
\end{equation}
where $q_{type}, type=j,l$ stand for the electric charge of baryons
and leptons respectively and
\begin{equation}
\mu_j=q_j \mu_n - q_e (\mu_e - \mu_{\nu}), \qquad \mu_\mu=\mu_e.
\label{beta}
\end{equation}
We have also used the non extensive statistics for the leptons,
  which enter the calculation as free particles obeying the above
  mentioned conditions.

The three snapshots of the time evolution of a neutron star in
its first minutes of life are given by:
\begin{itemize}
\item ${\cal S}/n_B=1$, $Y_l=0.3$,
\item ${\cal S}/n_B=2$, $\mu_{\nu}=0$,
\item ${\cal S}/n_B=0$, $\mu_{\nu}=0$,
\end{itemize}
where
\begin{eqnarray}
Y_l=\frac{\sum_l n_l}{n_B},
\end{eqnarray}
which, according to simulations \cite{Burrows:1986me}, can reach
$Y_l\simeq 0.3-0.4$. In the present work we are interested in finite
temperature systems and hence, most of the results refer to the first
two snapshots.

Another aspect of the evolution of compact stars that is worth
investigating is the direct Urca (DU) process, $n\to p+e^-+ \bar{\nu}_e$ \cite{urca}.
It is known that the cooling of the star by neutrino emission can
occur relatively fast if it is allowed, what happens when the proton fraction exceeds a critical
value $x_{\rm \scriptscriptstyle DU}$\cite{urca},  evaluated in terms of the leptonic
fraction as \cite{klaen06}:
\begin{equation}
x_{\rm \scriptscriptstyle DU} = \frac{1}{1 + (1 + x_e^{1/3})^3}  \:,
\label{durca}
\end{equation}
where $x_e = n_e/(n_e + n_{\mu})$ is the electron leptonic fraction, $n_e$ is the number
density of electrons and $n_{\mu}$ is the number density of muons.
Cooling rates of neutron stars seem to indicate that this fast cooling process does
not occur and, therefore, a constraint is set imposing that the direct Urca
process is only allowed in stars with a mass larger than $1.5 \, M_\odot$, or a
less restrictive limit,  $1.35 \, M_\odot$ \cite{klaen06}.
The DU process can also occur for hyperons, if they are taken into
account in the EOS. Although the
 neutrino luminosities in these processes  are much smaller than the ones
 obtained in the nucleon direct Urca process, they play an important role
 if they occur at densities below the nucleon direct Urca process
 \cite{urca2}. The process $\Lambda\to p + e + \bar \nu$, for
 instance,  may occur at densities below the nucleon DU onset. In the next
section we also investigate the effects of non extensive statistics on the onset of the DU process.

To make our results depend as little as possible
  on too many degrees of freedom, we start by analysing  the EOS  with
  nucleons only for different values of $q$. We also investigate
  the effects of non extensivity on a free Fermi gas at finite
  temperature, where nucleons and leptons are only subject to the
  conditions of $\beta$-equilibrium and charge neutrality.
As it is widely accepted that hyperons should be present inside
  (proto)neutron stars, they are also included and the effects of
  using different $q$ values, always larger than one, are checked.

\section{Results}
\label{results}

We now calculate and analyze stellar properties obtained with two
different values of the non extensive statistics $q$ parameter, namely
$q=1.05$ and 1.14. Our results are then compared with the ones shown
in Ref. \cite{lavagno}. We have chosen values larger than one because
lower values produce a slightly softer EOS, which result in lower
maximum stellar masses as compared with the standard non-linear
Walecka model, as can be seen in Ref. \cite{lavagno} and
therefore, may not useful if we want to explain massive compact
objects. In addition, in Ref.~\cite{Beck} it was shown that there is
an upper limit for the entropic index at $q_{max}=11/9$. On the other
hand, all experimental information on hadronic systems show that $q>1$.
Since our main goal is to check whether 2$M_\odot$ stars can be attained with the
help of non extensive thermodynamics when the traditional one fails,
we restrict ourselves to values that go in the desired direction.
In Ref. \cite{Lucas}, the entropic index $q$, is taken as a fixed
property of the hadronic matter with its value determined as $q=1.14$
from the analysis of $p_T$-distributions and in the study of the
hadronic mass spectrum. The value $q=1.05$ is used because it is
slightly larger than the value used in \cite{lavagno}, where the authors used
$q=1.03$ that represents just a small deviation from the standard
stellar matter physics. We have checked that the results obtained
  with $q=1.03$ and $q=1.05$ are numerically very similar. Had we
  plotted the next figures with both of them, the curves would be
  practically undistinguishable.

In all graphs shown next, the GM1 parametrization was used, but the
qualitative results are the same for the IU-FSU parameter set,
  despite the inclusion of the $\omega-\rho$ interaction.  Another
  aspect that we should mention is that quantum hadrodynamic models cannot
  describe the very low density part of the EOS well and they are usually
  linked to an appropriate EOS named BPS \cite{bps} at low
  densities and zero temperature. In the present work we have chosen
  not to use the BPS, which does not affect the macroscopic quantities
  we are interested in analyse. Moreover, non extensive thermodynamics
  is only valid at finite temperature, where the BPS would only be an
  approximate EOS.

\begin{figure}[tbp]
\centering
\subfloat[]{\includegraphics[width=0.68\linewidth,angle=270]{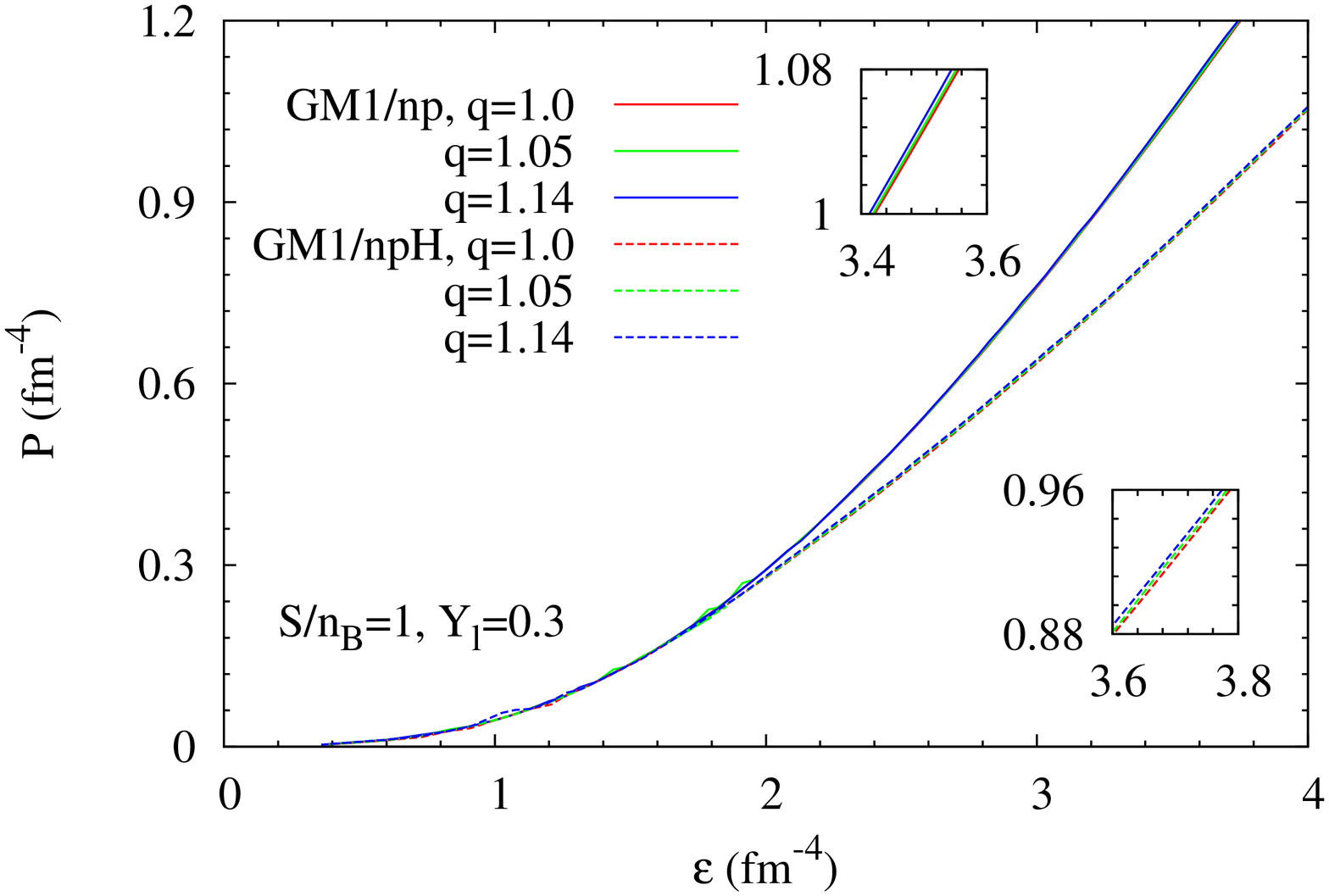}\label{fig1:a}}
\hfill
\subfloat[]{\includegraphics[width=0.68\linewidth,angle=270]{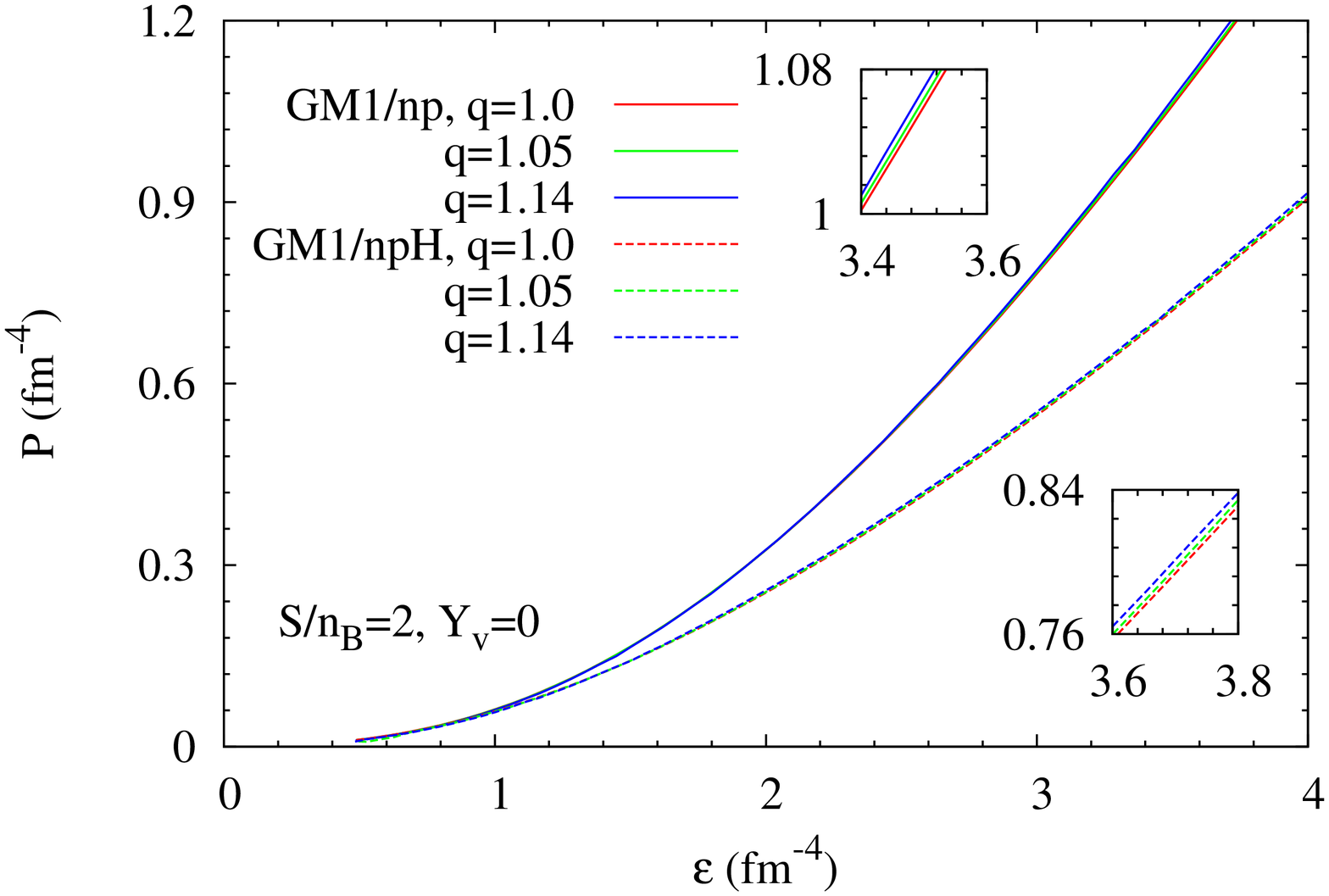}\label{fig1:b}}
\hfill
\subfloat[]{\includegraphics[width=0.68\linewidth,angle=270]{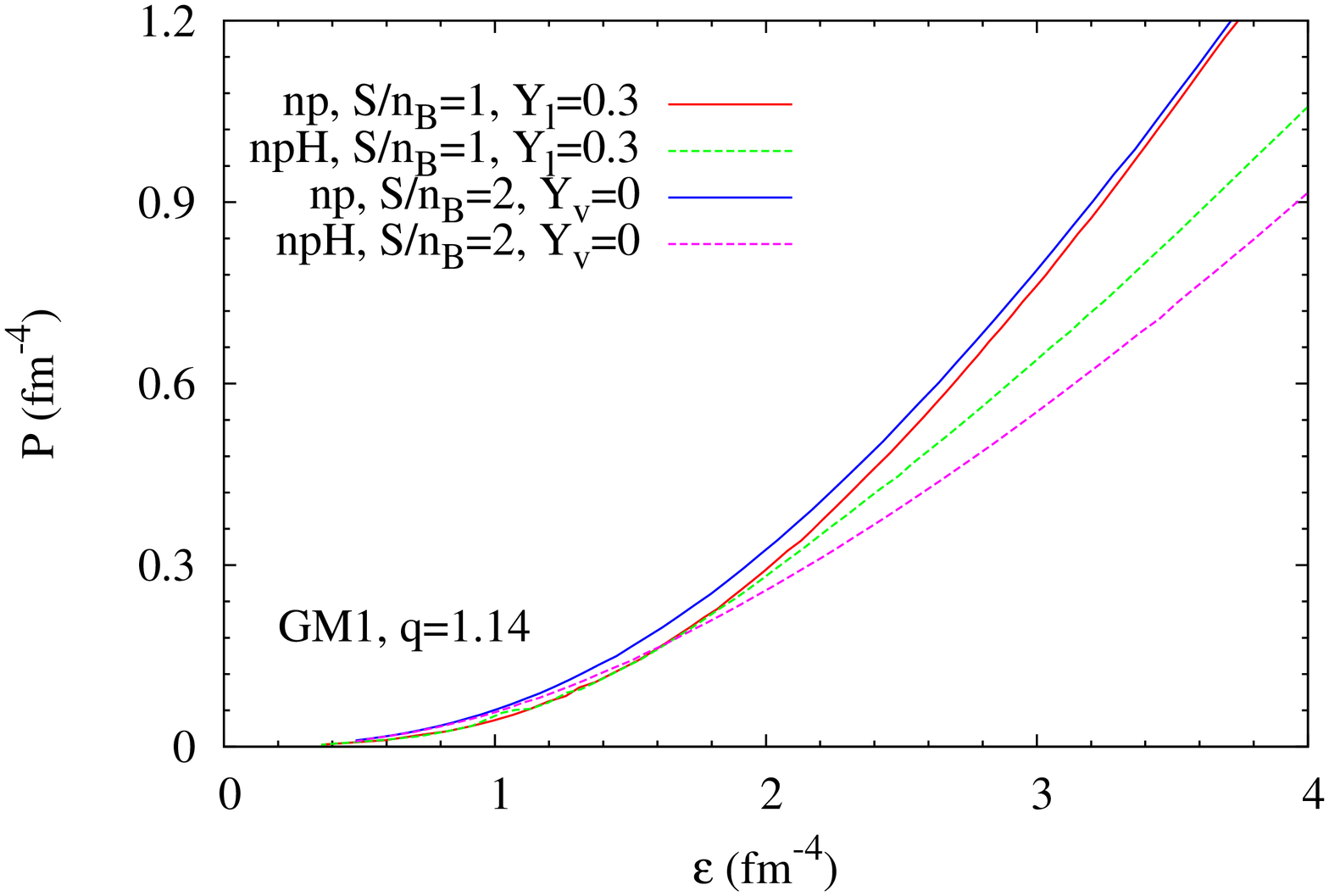}\label{fig1:c}}
\caption{Equation of state for hadronic matter constituted by
  nucleons only (np) and including the lightest eight baryons (npH)
  for different values of $q$ and a) first b) second snapshot of the
  star evolution c) for $q=1.14$ only and both snapshots.
\label{fig1} }
\end{figure}

We start by showing the EOS for the first two snapshots of the star
evolution in Fig.~\ref{fig1} for the cases with nucleons
only and also with hyperons. As it is always the case, hyperons
  make the EOS softer for a fixed $q$ value. It is difficult
  to distinghish the curves for our choice of $q$'s because numerically
  they are indeed very close, but not identical.
The deviation obtained with non extensive
statistics is very small, but larger at high densities for the
$q$-values we have considered, with consequences in the maximum
stellar masses, which will be seen later.
 It is important to observe that, for a fixed $q$-value, the EOS is slightly harder for
${\cal S}/n_B=2$, $\mu_{\nu}=0$ than for ${\cal S}/n_B=1$, $Y_l=0.3$
when only nucleons are taken as internal neutron star constituents,
but it is softer when the hyperons are considered. Nevertheless,
  this behaviour is valid also for the usual thermodynamics, when
  $q=1$ and hence, it is not a consequence of the use of non extensivity.

\begin{figure}[tbp]
\centering
\subfloat[]{\includegraphics[width=0.6\linewidth,angle=270]{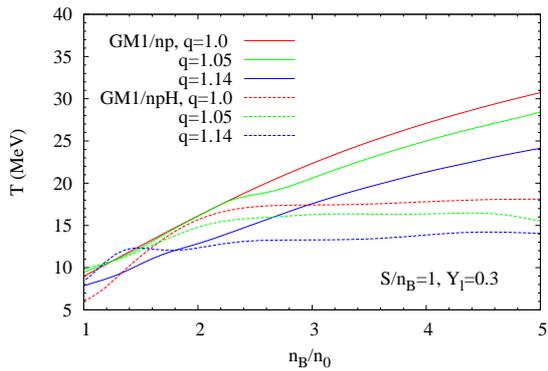}\label{fig2:a}}
\hfill
\subfloat[]{\includegraphics[width=0.6\linewidth,angle=270]{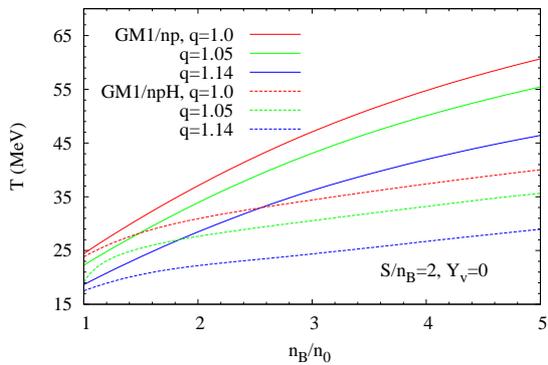}\label{fig2:b}}
\hfill
\subfloat[]{\includegraphics[width=0.6\linewidth,angle=270]{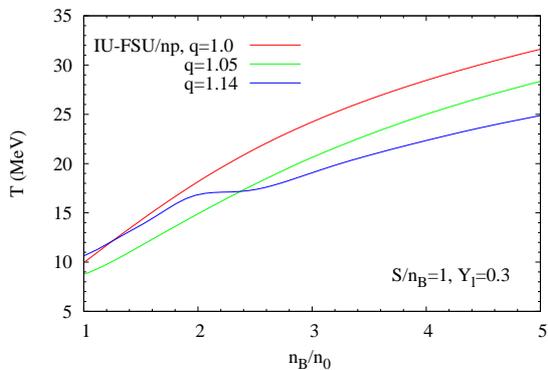}\label{fig2:c}}
\hfill
\subfloat[]{\includegraphics[width=0.6\linewidth,angle=270]{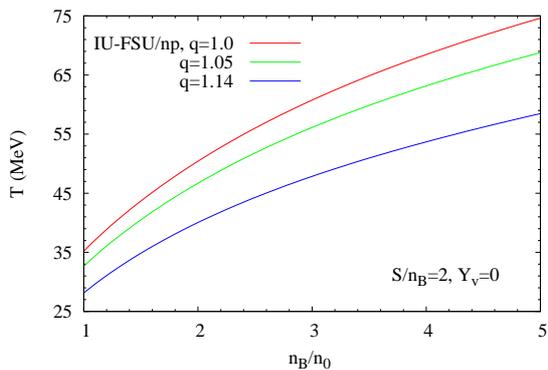}\label{fig2:d}}
\caption{Temperature as a function of density (in units of nuclear
  matter saturation density) for  different values of $q$ and a) GM1
  and first b) GM1 and second c) IU-FSU and first and D)
  IU-FSU and second snapshot of the star evolution. \label{fig2} }
\end{figure}

We then analyze the effects of non extensivity on the internal stellar
temperature by plotting the temperature as a function of density again
for the first two snapshots of the star evolution in Fig. \ref{fig2}
for both parametrizations investigated in the present work.
 We display temperature results for densities higher than nuclear
  matter saturation density because at subsaturation densities, the
  EOS would be more similar to the one of a free Fermi gas and at very low
  densities a  BPS-like EOS would have to be employed, what we have not
  done. Nevertheless, we can see from Tables \ref{star-properties-gm1} and \ref{star-properties-iufsu} that the
  effects of non extensivity on a free gas at fixed temperature are
  very small, which means that the curves would tend to get closer to
  each other as the density and the temperature decrease.
 Then, we clearly see that
the temperature decreases with the increase of $q$, a behavior already
expected from the calculations performed in \cite{ours} (see, for
instance figs. 2 and 6 of that reference). At densities of the order
of 5 times nuclear saturation density, the temperature decreases by
approximately 25\% in average, with important consequences in the neutrino
diffusion during the Kevin-Helmholtz epoch, when the star evolves from
a hot and lepton rich object to a cold and deleptonized compact star.
The cooling would be faster for larger $q$ values.
 However, in Ref. \cite{lavagno}, the behavior is exactly the
 opposite, i.e. the temperature increases with the increase of the
 $q$- value, a result that we do not reproduce.  We do not believe
   that the use of a different expression for the partition function,
   as we have commented in section \ref{subsec:nonextensive} is responsible for this
   opposite behaviour.

\begin{figure}[tbp]
\centering
\subfloat[]{\includegraphics[width=0.68\linewidth,angle=270]{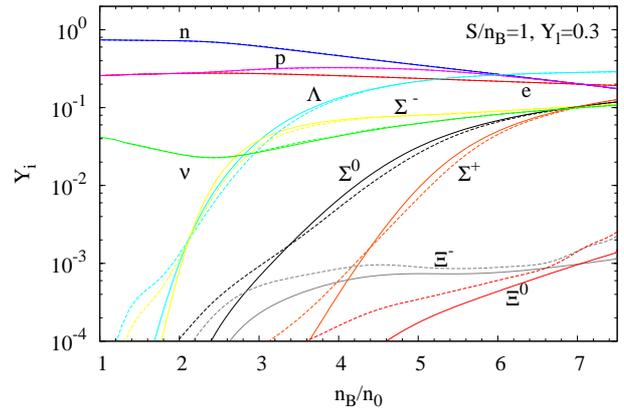}\label{fig3:a}}
\hfill
\subfloat[]{\includegraphics[width=0.68\linewidth,angle=270]{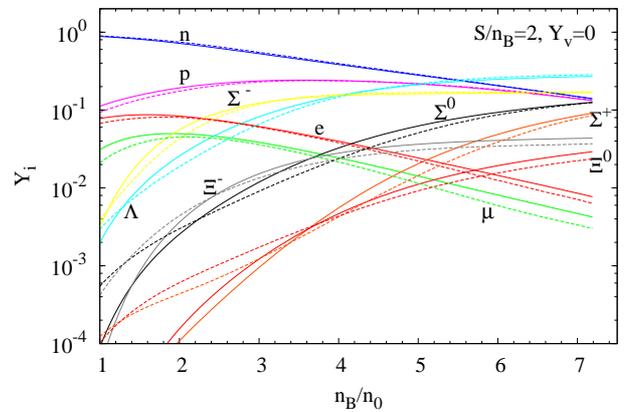}\label{fig3:b}}
\caption{ Particle fractions obtained for a) first and  b) second snapshot of the
  star evolution.
We use solid line for $q=1.0$ (standard model) and
  dashed lines for $q=1.14$.
 \label{fig3}}
\end{figure}

In order to see how the internal constitution of the star  is affected by
non extensivity, we plot in Fig. \ref{fig3}  the particle fractions
when the hyperons are considered and the related strangeness content
in Fig. \ref{fig4}. We do not include the particle concentrations
  for the case with nucleons only because, they are affected very
  little by non extensivity, as shown in Ref.~\cite{lavagno}.
From the figures we plot, we can see that as $q$ increases, the
amount of strangeness decreases, which means that the EOS becomes
harder, resulting in larger maximum masses. On the other hand, we
  already knew that in a free system, heavier particles are favored
  when $q$ becomes larger than one, as seen in Fig. \ref{fig00}, a
  behaviour that is also observed in Fig. \ref{fig3}, i.e., the onset
  of hyperons takes place at lower densities. But stellar matter is
  also subject to charge neutrality, $\beta$-equilibrium and
different values of temperature at different densities and the
  final balance results in a system with a slightly smaller 
  strangeness content for larger $q$ values. The numbers used in
    Fig \ref{fig4} for the case  $S/n_B=2$ at $n_B/n_0 \simeq 3$
tell us that as $q$ goes from 1 to 1.05, the decrease in the strangeness
content is 2.5\%, when it goes from 1.05 to 1.14, it reaches 4.7\% and
from $q=1$ to $q=1.14$, the decrease is of the order of 7.1\%. This
decrease makes the EOS harder and hence the explanation for the
slightly larger maximum masses obtained with non extensivity.

\begin{figure}[tbp]
\centering
\subfloat[]{\includegraphics[width=0.68\linewidth,angle=270]{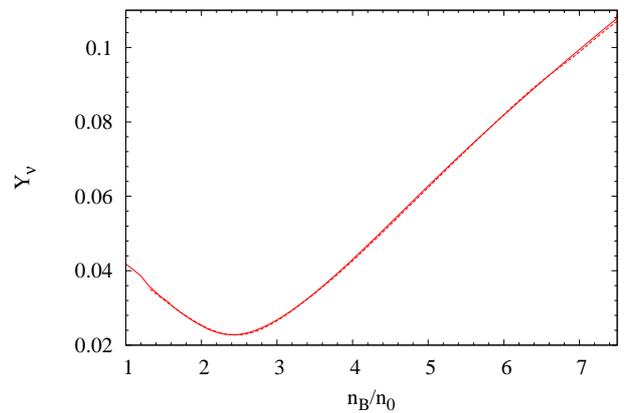}}
\caption{Neutrino content obtained for the first snapshot of the star
  evolution. We use solid lines for $q=1.0$ (standard model) and
  dashed lines for $q=1.14$.
\label{fignew}}
\end{figure}

\begin{figure}[tbp]
\centering
\includegraphics[width=0.68\linewidth,angle=270]{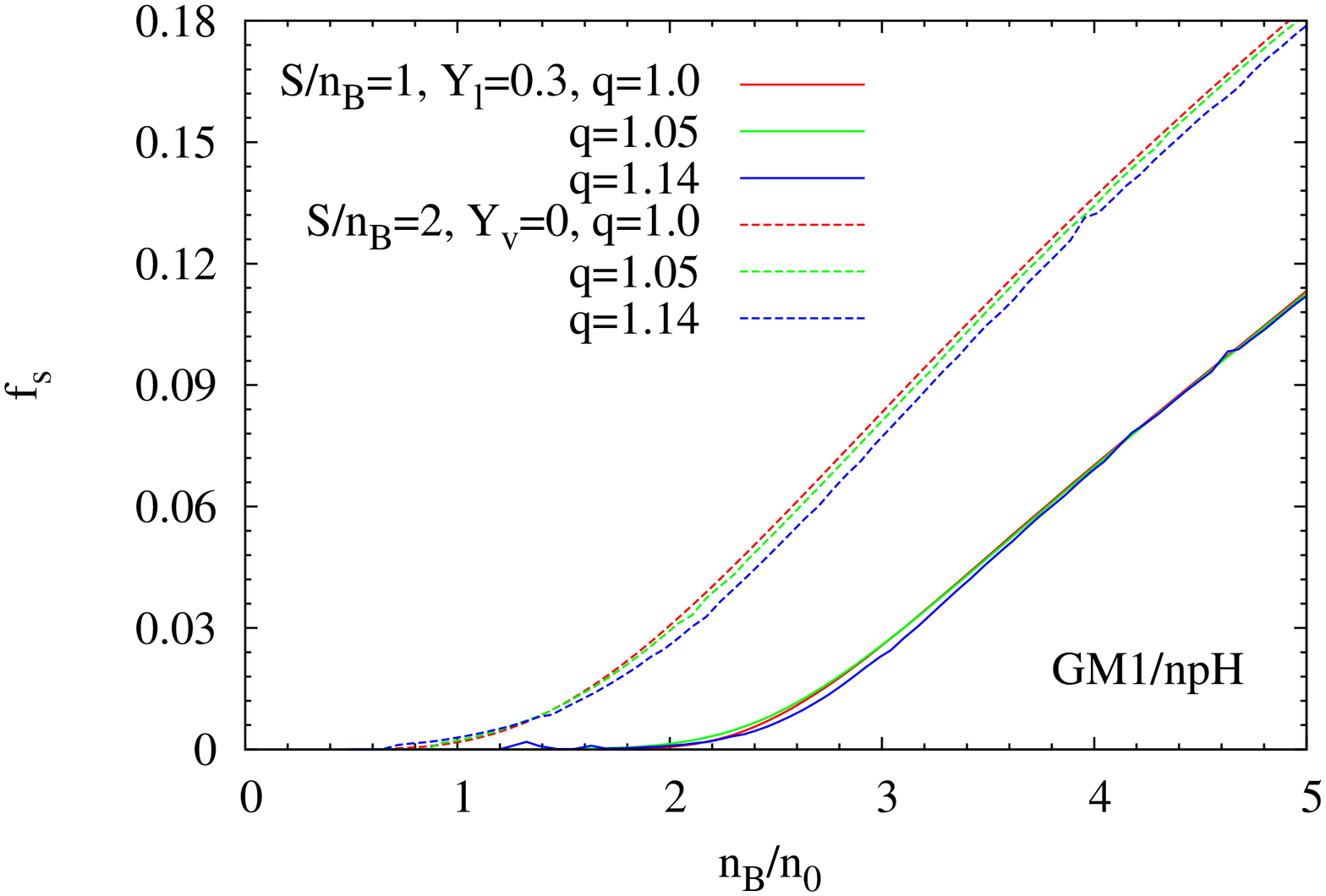}\\
\caption{Strangeness content as a function of density  (in units of nuclear
  matter saturation density) for  different values of $q$ and two first snapshots of the
  star evolution. \label{fig4}}
\end{figure}

Neutrinos also play an important role when the lepton fraction is
  fixed, during the first snapshot of the star evolution. If hyperons
  are included, neutrinos help in making the EOS harder, but affect
  very little the EOS if only nucleons are present in the system. This
  is a well known result for $q=1$. In Fig. \ref{fignew} we plot only
  the neutrino fraction, so that its behaviour with $q$ becomes evident.
Non extensivity practically does not change the amount of
  neutrinos.
In the difusion approximation normally used in the calculation of the
temporal evolution of protoneutron stars in the Kelvin-Helmholtz
phase, the neutrino  mean free path depends on the diffusion
coefficientes,  obtained from the EOS and dependent of the neutrino
fraction and distribution function. Hence, any change in the
  neutrino content would certainly influence the stellar evolution,
  but non extensivity seems not to affect this quantity in a
  non-neglectable way.

\begin{figure}[tbp]
\centering
\subfloat[]{\includegraphics[width=0.68\linewidth,angle=270]{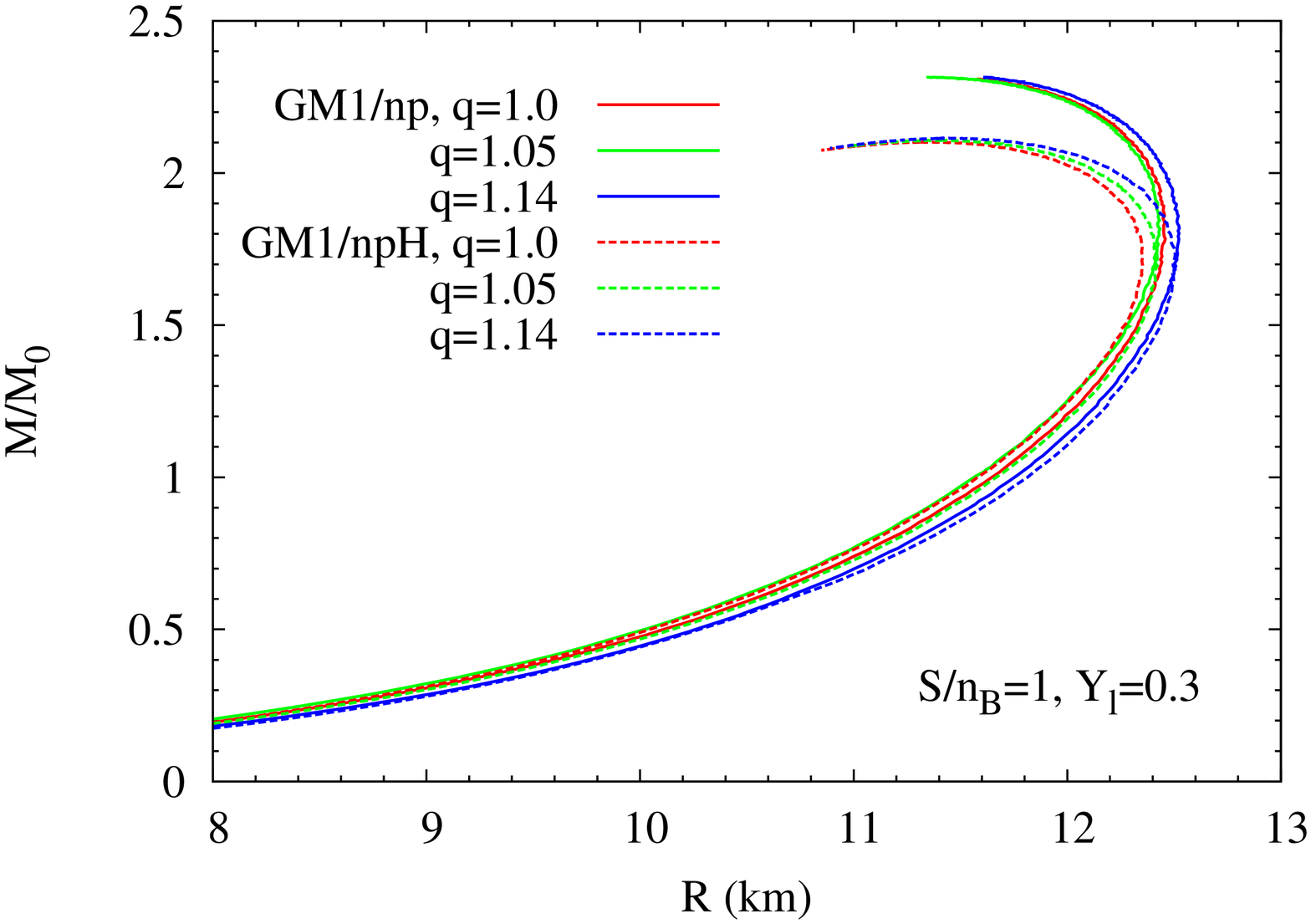}\label{fig5:a}}
\hfill
\subfloat[]{\includegraphics[width=0.68\linewidth,angle=270]{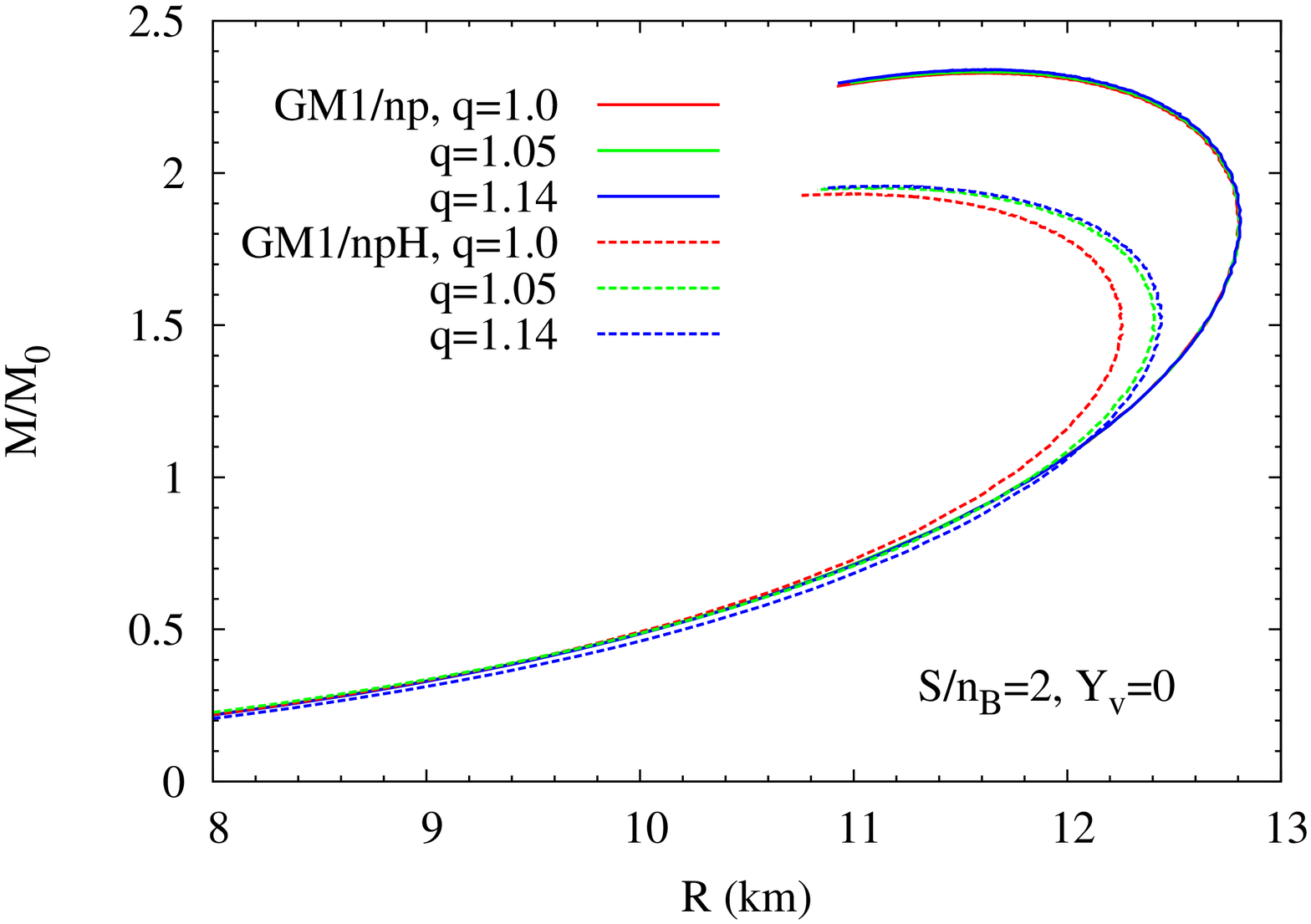}\label{fig5:b}}
\hfill
\subfloat[]{\includegraphics[width=0.68\linewidth,angle=270]{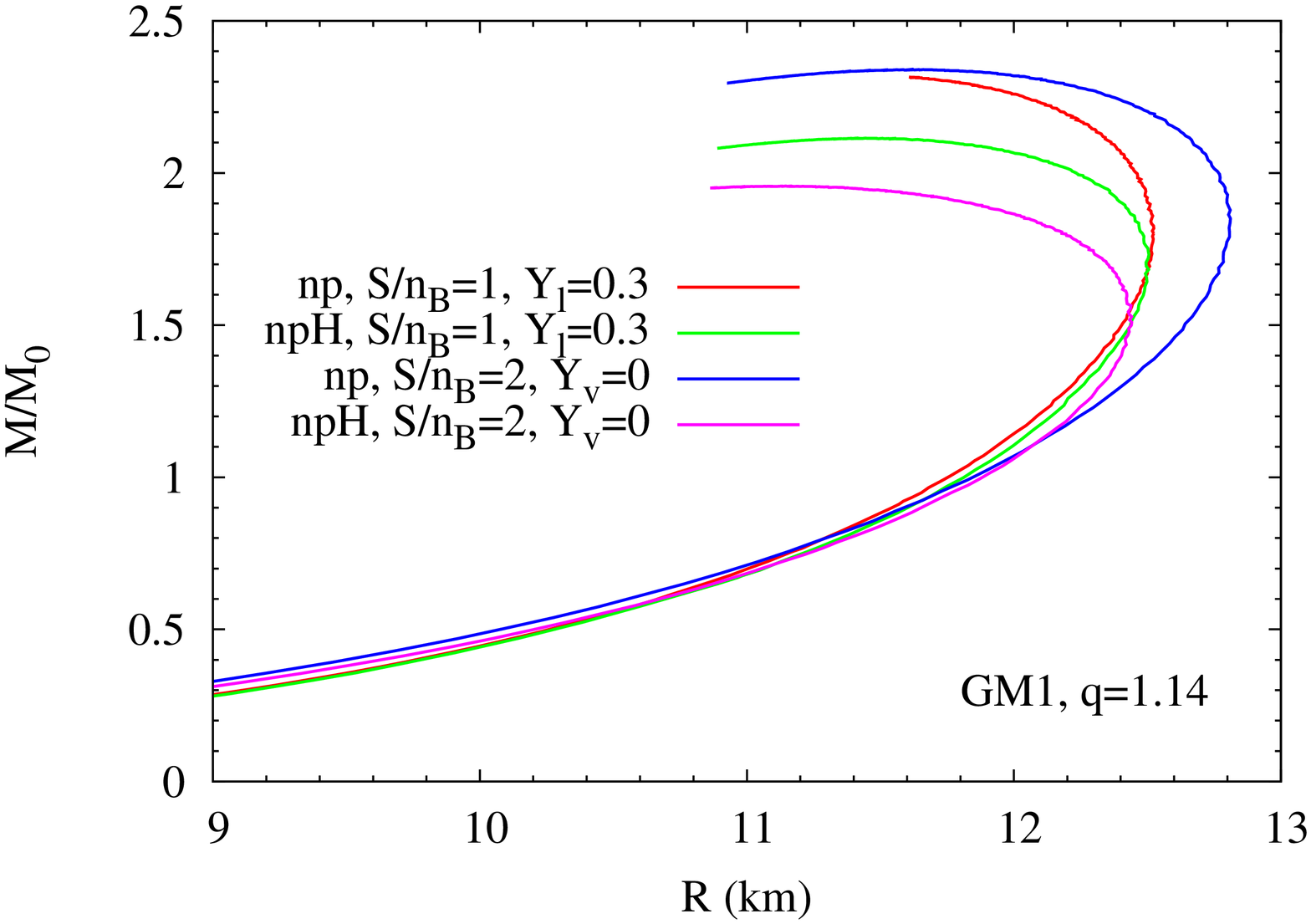}\label{fig7:b}}
\caption{Mass-radius results obtained from the solution of the TOV
  equations for hadronic matter constituted by
  nucleons only (np) and including the lightest eight baryons (npH)
  for different values of $q$ and a) first b) second snapshot of the
  star evolution c) for $q=1.14$ only and both snapshots.
\label{fig5} }
\end{figure}

In Table \ref{star-properties-gm1} and Fig. \ref{fig5} we show the main stellar
properties obtained from the solution of the Tolman-Oppenheimer-Volkof
(TOV) equations \cite{tov}, which use the EOS just discussed with the
GM1 parametrization as input. As expected, from the observation of the
EOS, the maximum stellar mass increases with the increase of the $q$-
value. When only nucleons are taken into account, the maximum stellar
masses are obtained during the second snapshot of the star evolution
(${\cal S}/n_B=2$, $\mu_{\nu}=0$) and when hyperons are also included,
maximum masses come out for ${\cal S}/n_B=1$, $Y_l=0.3$.  This
behavior corroborates the findings in Ref.~\cite{lavagno}. For the
sake of completeness, we also display results obtained at fixed
temperature ($T=30$ MeV) and compare them with the results for a free
Fermi gas, in which case neutrons, protons, electrons and muons
  obey stellar matter conditions, but are not subject to nuclear
  interaction.
This temperature was chosen to be of interest in the stellar
  medium, durind the cooling process, as seen in Fig. \ref{fig4}. Of
  course, we could have chosen $T=20$ MeV instead, a temperature at
  which chemical freeze-out takes place in heavy ion collisions, but
  the numerical results would be very similar.
 In the cases where GM1 was used, there is no obvious
pattern with respect to the $q$-value, i.e. the maximum masses
oscillate when the $q$-value increases. When a free Fermi gas is used,
the maximum masses decrease when $q$ increases.  As it is well known
the huge increase in the maximum masses is due to the inclusion of the
nuclear interaction, but we also found a lack of pattern in a system
with fixed temperature instead of fixed entropy.  In Fig. \ref{fig5}
we plot mass-radius results obtained from the EOS shown in
Fig. \ref{fig1}. In these curves the BPS \cite{bps} EOS was not
included because it is only valid at zero temperature and, as shown in
Fig. \ref{fig2}, the temperature at the surface of the star for fixed
entropies can be slightly higher. Had we included the BPS EOS, our
curves would present a tail towards higher radii, but the differences
in the maximum masses would be minor.

To check the consistency of our results, in Table \ref{star-properties-iufsu} we display, for a system with
nucleons only, stellar properties obtained with the IU-FSU
parametrization. We have not included hyperons because this parameter
sets provides too low maximum stellar masses when strangeness is taken
into account. The results show that the qualitative conclusions
  with respect to the effects of non extensivity do not
depend on the chosen parameter set, even when extra crossing terms
  involving the $\omega-\rho$ interaction is considered.

\begin{table*}[ht]
\centering
\begin{tabular}{ccccccc}
\hline
model & case & $q$ & $M_{max}$ & $Mb_{max}$ & R & ${\cal E}_0$ \\
 &&& ($M_\odot$) & ($M_\odot$) & (Km) & (fm$^{-4}$) \\
\hline
free gas & T=30 MeV, $Y_{\nu}=0$ & 1.0   & 0.693 & 0.70 & 7.46 & 12.59 \\
free gas & T=30 MeV, $Y_{\nu}=0$& 1.05 & 0.689 & 0.70 & 7.30 & 13.48 \\
free gas & T=30 MeV, $Y_{\nu}=0$& 1.14 & 0.680 & 0.69 & 7.06 & 14.32 \\
\hline
\hline
GM1/np & T=30 MeV, $Y_{\nu}=0$& 1.0 & 2.10 & 2.37 & 11.48 & 5.83 \\
GM1/np & T=30 MeV, $Y_{\nu}=0$& 1.05 & 2.30 & 2.66 & 11.34 & 6.00 \\
GM1/np & T=30 MeV, $Y_{\nu}=0$& 1.14 & 2.29 & 2.64 & 11.36 & 5.85 \\
\hline
GM1/np & ${\cal S}/n_B=1$, $Y_l=0.3$ & 1.0 & 2.31 & 2.67 & 11.57 & 5.18 \\
GM1/np & ${\cal S}/n_B=1$, $Y_l=0.3$ & 1.05 & 2.31 & 2.67 & 11.38 & 5.77 \\
GM1/np & ${\cal S}/n_B=1$, $Y_l=0.3$ & 1.14 & 2.32 & 2.68 & 11.61 & 5.18 \\
\hline
GM1/np & ${\cal S}/n_B=2$, $Y_{\nu}=0$& 1.0 & 2.33 & 2.66 & 11.60 & 5.71 \\
GM1/np & ${\cal S}/n_B=2$, $Y_{\nu}=0$& 1.05 & 2.33 & 2.68 & 11.64 & 5.62 \\
GM1/np & ${\cal S}/n_B=2$, $Y_{\nu}=0$& 1.14 & 2.34 & 2.70 & 11.61 & 5.71 \\
\hline
GM1/np & T=0, $Y_{\nu}=0$ & 1.0 & 2.38 & 2.88 & 11.75 & 5.62 \\
\hline
\hline
GM1/npH & T=30 MeV, $Y_{\nu}=0$ & 1.0 & 1.90 & 2.12 & 10.88 & 6.28 \\
GM1/npH & T=30 MeV, $Y_{\nu}=0$ & 1.05 & 1.90 & 2.11 & 10.73 & 6.78\\
GM1/npH & T=30 MeV, $Y_{\nu}=0$ & 1.14 & 1.89 & 2.07 & 10.61 & 6.93\\
\hline
GM1/npH & ${\cal S}/n_B=1$, $Y_l=0.3$ & 1.0 & 2.10 & 2.39 & 11.40 & 5.69 \\
GM1/npH & ${\cal S}/n_B=1$, $Y_l=0.3$ & 1.05 & 2.11 & 2.54 & 11.43 & 5.72 \\
GM1/npH & ${\cal S}/n_B=1$, $Y_l=0.3$ & 1.14 & 2.11 & 2.39 & 11.44 & 5.84 \\
\hline
GM1/npH & ${\cal S}/n_B=2$, $Y_{\nu}=0$& 1.0 & 1.93 & 2.15 & 10.98 & 6.46 \\
GM1/npH & ${\cal S}/n_B=2$, $Y_{\nu}=0$& 1.05 & 1.95 & 2.18 & 11.10 & 6.29 \\
GM1/npH & ${\cal S}/n_B=2$, $Y_{\nu}=0$& 1.14 & 1.96 & 2.20 & 11.13 & 6.26 \\
\hline
GM1/npH & T=0, $Y_{\nu}=0$ & 1.0 &  2.00  & 2.32 & 11.51 &  5.96 \\
\hline
\hline
\end{tabular}
\caption{Protoneutron star macroscopic properties (maximum gravitation mass, maximum baryonic mass, corresponding radius and central energy density) for different values of $q$ and fixed temperature or one of the snapshots of the star evolution.}
\label{star-properties-gm1}
\end{table*}

\begin{table*}[ht]
\centering
\begin{tabular}{ccccccc}
\hline
model & case & $q$ & $M_{max}$ & $Mb_{max}$ & R & ${\cal E}_0$ \\
 &&& ($M_\odot$) & ($M_\odot$) & (Km) & (fm$^{-4}$) \\
\hline
free gas & T=30 MeV, $Y_{\nu}=0$ & 1.0   & 0.693 & 0.70 & 7.46 & 12.59 \\
free gas & T=30 MeV, $Y_{\nu}=0$& 1.05 & 0.689 & 0.70 & 7.30 & 13.48 \\
free gas & T=30 MeV, $Y_{\nu}=0$& 1.14 & 0.680 & 0.69 & 7.06 & 14.32 \\
\hline
\hline
IU-FSU/np & T=30 MeV, $Y_{\nu}=0$& 1.0 & 1.90 & 2.19 & 10.76 & 6.08 \\
IU-FSU/np & T=30 MeV, $Y_{\nu}=0$& 1.05 & 1.90 & 2.19 & 10.71 & 6.46 \\
IU-FSU/np & T=30 MeV, $Y_{\nu}=0$& 1.14 & 1.89 & 2.16 & 10.57 & 6.84 \\
\hline
IU-FSU/np & ${\cal S}/n_B=1$, $Y_l=0.3$ & 1.0 & 1.89 & 2.13  & 10.55 &6.59\\
IU-FSU/np & ${\cal S}/n_B=1$, $Y_l=0.3$ & 1.05 & 1.94 & 2.25 & 11.33 & 6.34 \\
IU-FSU/np & ${\cal S}/n_B=1$, $Y_l=0.3$ & 1.14 & 1.94 & 2.14 & 11.30 & 6.34 \\
\hline
IU-FSU/np & ${\cal S}/n_B=2$, $Y_{\nu}=0$& 1.0 & 1.97 & 2.19 & 11.29 & 5.86 \\
IU-FSU/np & ${\cal S}/n_B=2$, $Y_{\nu}=0$& 1.05 & 1.97 & 2.20 & 11.25 & 5.96 \\
IU-FSU/np & ${\cal S}/n_B=2$, $Y_{\nu}=0$& 1.14 & 1.98 & 2.22 & 11.24 & 5.97 \\
\hline
IU-FSU/np & T=0, $Y_{\nu}=0$ & 1.0 & 1.95 & 2.28 & 10.82 & 6.37 \\
IU-FSU/npH & T=0, $Y_{\nu}=0$ & 1.0 &  1.52  & 1.71 & 10.31 &  6.90 \\
\hline
\hline
\end{tabular}
\caption{Same as Table \ref{star-properties-gm1}}
\label{star-properties-iufsu}
\end{table*}

We now analyse the radii results.  According to Ref.~\cite{Hebeler},
the radii of the canonical $1.4M_\odot$ neutron star should lie in the range
9.7-13.9 Km. Based on an analysis in which it was assumed that all
neutron stars have the same radii, they should lie in the
range $R=9.1^{+1,3}_{-1.5}$ \cite{guillot} and another calculation, based
on a Bayesian analysis, foresees radii of all neutron stars to lie in
between 10 and 13.1 Km~\cite{Lattimer2013}.
The radii results shown in Tables \ref{star-properties-gm1} and \ref{star-properties-iufsu} correspond to the maximum
  mass stars. If only nucleons are considered as
  neutron star constituents, as $q$ increases no general pattern is
 found for the resulting radii. However, when hyperons are taken into
account, the radii increase with the increase of $q$. These radii,
even for $q=1.14$ are not too large, varying around 11.5 Km.
However, if we consider the
radii of the canonical 1.4 M$_\odot$ stars, we can see, from Fig.
\ref{fig5} that non extensivity generally makes them slightly larger and they
stand around 12 Km, a somewhat large value if  the above mentioned constraints
are to be taken seriously. Had we included the BPS EOS, they would be
still a bit larger. Let's stress that the radii are determined by the
parametrization chosen, depending also on the hyperon-meson coupling
constants. Hence, if a model succeeds in describing a small radius,
non extensivity is not likely to modify it too much.

\begin{figure}[tbp]
\centering
\subfloat[]{\includegraphics[width=0.68\linewidth,angle=270]{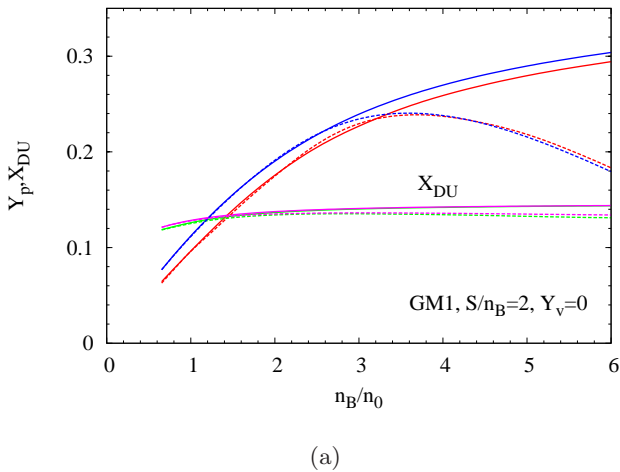}}
\caption{Onset of direct Urca process in stellar matter with nucleons
  only (solid lines) and with hyperons (dashed lines) for the usual Boltzmann-Gibbs statistics
  (q=1, blue and pink lines) and for $q=1.14$ (red and green lines).
\label{fig6} }
\end{figure}

Finally, in Fig. \ref{fig6} we plot the onset of the direct Urca
process in stellar matter for matter with (dashed lines) and without
hyperons (solid lines) in the case where ${\cal S}/n_B=2$, $\mu_{\nu}=0$.
The lines around a y-value of 0.12 refer to $x_{DU}$ and the other
lines represent the proton fraction. When the curves cross, we can see
the value of the proton fraction and the respective baryonic density.
We can see that the line for $x_{DU}$ coincides for the standard model
independently of considering or not hyperons. For $q=1.14$ both curves
present a small deviation at large densities.
For GM1, the standard density value for which the DU process occurs
(at zero temperature and matter without hyperons) is 1.81 times
nuclear matter saturation density \cite{rafael11} .
When we fix the entropy density to 2 and keep $q=1$, this value
decreases to 1.207 (1.205) $n_B/n_0$ with (without) hyperons
but when we look at the values for $q=1.14$, we see that the onset of
the DU process increases again by approximately 21.5\% to  1.423
(1.402) $n_B/n_0$ with (without) hyperons. The proton fraction that we
obtain with nucleons only and with hyperons are coincident for a fixed $q$-value at low densities
and just deviate from each other when other hyperons with positive charge
appear.  Therefore, if the DU process determines how the star cools
down, a system described by non extensive statistics would certainly affect the
  interpretation of the cooling rate mechanism.

\section{Final Remarks}
\label{conclusion}

We have applied non extensive statistics to calculate equations of state
that describe stellar matter with two of the commonly used
parametrizations for the non-linear Walecka model, namely GM1
\cite{gm1} and IU-FSU \cite{iufsu}. We have then fixed two $q$-values
(1.05 and 1.14) and obtained the most important microscopic
quantities associated with the equations of state, i.e.
particle fractions, strangeness, internal temperature and direct Urca
process onset for two snapshots of the star evolution.
As compared with the existing work on the application of non
  extensive thermodynamics \cite{lavagno}, some new features were
  investigated, apart from the use of two different parameter sets.
We have confirmed that the equations of state are only slightly modified, but
the effects are enough to produce stars with slightly higher maximum
masses and these results are common to both parameter sets used.
However, contrary to what was obtained in \cite{lavagno}, we
  found that the internal temperature of the stars decreases with the increase of the
$q$-value  and at densities of the order
of 5 times nuclear saturation density, the temperature decreases by
approximately 25\% in average, with important consequences in the neutrino
diffusion during the Kevin-Helmholtz epoch, when the star evolves from
a hot and lepton rich object to a cold and deleptonized compact
star. This aspect should certainly be better investigated.
Moreover, we have also seen that the direct Urca process is substantially affected
by non extensivity, with consequences on the cooling rates of the stars.

As usually done in the search for macroscopic star properties, the
Tolman-Oppenheimer-Volkof equations were then solved for the
previously obtained EOS and the macroscopic quantities were computed.
The results were compared with more academic calculations for fixed
temperatures and for a free-Fermi gas but in these cases, due to
  the imposition of stellar matter constraints, no specific pattern
  was found, differently from what happens in a system with really
  free gases, as seen in Figs. \ref{fig0}, \ref{fig00} and in
  Ref.~\cite{ours}.

A final word on our choice of $q$ values larger than one is
  worthy. All experimental information on hadronic systems show that
  $q>1$ and according to Ref.~\cite{Beck}, there is
an upper limit for the entropic index at $q_{max}=11/9$.
We have checked that it is possible to use the sub-extensive regime
with the help of an appropriate expansion, but it does not make sense
in applications to protoneutron stars if the desired effect is to
increase the maximum mass and this regime makes it decrease.

In this work, we have also used the non extensive statistics for the leptons,
  which enter the calculation as free particles with respect to the
  strong nuclear interaction, but subject to the conditions of charge
  neutrality and $\beta$-equilibrium. We could have used different
  $q$-values for the leptons, but for simplicity, we have opted to use
  the same values as for the baryons.

\begin{acknowledgement}
This work was partially supported by CNPq (grants 305639/2010-2,
300602/2009-0, 455719/2014-4 and 304105/2014-7), FAPESC (Brazil) under
project 2716/2012,TR 2012000344, FAPESP (Brazil) under grant
2013/24468-1, Spanish Plan Nacional de Altas Energ\'{\i}as grant
FPA2011-25948, Junta de Andaluc{\'\i}a grant FQM-225, Generalitat de
Catalunya grant 2014-SGR-1450, Spanish MINECO's Consolider-Ingenio
2010 Programme CPAN (CSD2007-00042), Centro de Excelencia Severo Ochoa
Programme grant SEV-2012-0234. The research of E.M. has been supported by the
Juan de la Cierva Program of the Spanish MINECO, and by the European
Union under a Marie Curie Intra-European Fellowship (FP7-PEOPLE-2013-IEF) with project number PIEF-GA-2013-623006.
\end{acknowledgement}


\begin{thebibliography}{}
%
\bibitem{Lattimer2004} J. M. Lattimer and M. Prakash, Science {\bf 304},
536 (2004).

\bibitem{keil1995} W. Keil and H. Janka, Astron. Astrophys. {\bf 296}, 145
(1995).

\bibitem{pons} J. A. Pons, S. Reddy, M. Prakash, J. M. Lattimer and J.A.
Miralles, Astrophys. J. 513, 780 (1999); J. A. Pons, A. W. Steiner, M. Prakash and J. M. Lattimer,
Phys. Rev. Lett. {\bf 86} (2001) 5223-5226.

\bibitem{Prakash:1996xs}
  M.~Prakash, I.~Bombaci, M.~Prakash, P.~J.~Ellis, J.~M.~Lattimer and R.~Knorren,
  Phys.\ Rept.\  {\bf 280}, 1 (1997).

\bibitem{sw} B.D. Serot and J.D. Walecka, Adv. Nucl. Phys. {\bf 16}
(1986) 1.; J. Boguta and A. R. Bodmer, Nucl. Phys. {\bf A 292}, 413 (1977).

\bibitem{glen00} N.~K.~Glendenning, {\it Compact Stars}, Springer-Verlag, New-York, 2000.

\bibitem{Haensel}P.~Haensel, A.~Y.~Potekhin, D.~G.~Yakovlev:
{\it Neutron Stars, Equation of State and Structure},
Springer, New York  (2006).

\bibitem{demorest} Paul Demorest, Tim Pennucci, Scott Ransom, Mallory
  Roberts, and Jason Hessels, Nature (London) 467, 1081 (2010).

\bibitem{antoniadis} J. Antoniadis et al, Science 26, 340 n. 6131 (2013).

\bibitem{Tsallis88} C. Tsallis, J. Stat. Phys. 52 (1988) 479.

\bibitem{TsallisBook} M.~Gell-Mann, C.~Tsallis, {\it Nonextensive Entropy: Interdisciplinary Applications}, Oxfor University Press, USA, 2004. 

\bibitem{Bediaga} I. Bediaga, E.M.F. Curado e J.M. de Miranda,
Physica A 286 (2000) 156.

\bibitem{Deppman12} A. Deppman, Physica A 391 (2012) 6380; Physica A
  400 (2014) 207.
  
\bibitem{lavagno} A. Lavagno and D. Pigato, Eur. Phys. J. A (2011)
  {\bf 47}: 52.
  
\bibitem{Beck} C. Beck, Physica A 286, 164 (2000).

\bibitem{CW} J. Cleymans e D. Worku, J. Phys. G: Nucl. Part. Phys. 39 (2012) 025006.

\bibitem{Hagedorn} R.~Hagedorn, Lect.\ Notes Phys.\  {\bf 221}, 53 (1985).

\bibitem{Lucas} L. Marques, E. Andrade-II and A. Deppman, Phys. Rev. D
  87, 114022 (2013).

\bibitem{Lucas2} 
  L.~Marques, J.~Cleymans and A.~Deppman,  Phys.\ Rev.\ D {\bf 91}, 054025 (2015).

\bibitem{Sena} I. Sena and A. Deppman, Eur. Phys. J. A 49 (2013) 17.

\bibitem{Deppman2} 
  A.~Deppman,  J.\ Phys.\ G {\bf 41}, 055108 (2014).
  
\bibitem{ours} Eugenio Meg\'ias, D\'ebora P. Menezes and Airton
  Deppman, Physica A: Statistical Mechanics and its Applications 421, 15-24
(2015).

\bibitem{Megias:2014tha}
  E.~Meg\'ias, D.~P.~Menezes and A.~Deppman, EPJ Web Conf.\  {\bf 80}, 00040 (2014).

\bibitem{Deppman:2015cda} 
  A.~Deppman, E.~Meg\'ias and D.~Menezes, J.\ Phys.\ Conf.\ Ser.\  {\bf 607}, no. 1, 012007 (2015).

\bibitem{jirina}  M. Dutra, O. Louren\c co, S. S. Avancini, B. v. Carlson,
A. Delfino, D. P. Menezes, C. Providencia, S. Typel and J. R. Stone,
Phys. Rev. C 90, 055203 (2014).

\bibitem{gm1} N. K. Glendenning and S. A. Moszkowski,
Phys. Rev. Lett. {\bf 67}, 2414 (1991)

\bibitem{iufsu} F. J. Fattoyev, C. J. Horowitz, J. Piekarewicz and G. Shen,
Phys. Rev. C {\bf 82}, 055803 (2010); (arxiv:1008.3030).

\bibitem{hor01} C. J. Horowitz and J. Piekarewicz,
Phys. Rev. Lett. {\bf 86}, 5647 (2001).

\bibitem{fsu} B. G. Todd-Rutel and J. Piekarewicz, Phys. Rev. Lett. {\bf 95}, 122501 (2005);
F. J. Fattoyev and J. Piekarewicz, Phys. Rev. C {\bf 82}, 025805 (2010).

\bibitem{cp2014} Constan\c ca Provid\^encia, Sidney
S. Avancini, Rafael Cavagnoli, Silvia Chiacchiera, Camille Ducoin,
Fabrizio Grill, Jerome Margueron, D\'ebora P. Menezes, Aziz Rabhi, Isaac
Vidaña,  Eur. Phys. J. A (2014) 50:44.

\bibitem{hyp1} J. Schaffner-Bielich, A. Gal, Phys. Rev. C 62, 034311 (2000);
J. Schaffner-Bielich, M. Hanauske, H. Stocker and W. Greiner, Phys. Rev. Lett. 89, 171101 (2002);
E. Friedman, A. Gal, Phys. Rep. 452, 89 (2007).

\bibitem{lopes2014} Luiz L. Lopes and Debora P. Menezes,
Phys. Rev. C 89, 025805 (2014).

\bibitem{Plastino} J. M. Conroy, H. G. Miller and A. R. Plastino, Physics
  Letters A 374 (2010) 4581–4584.

\bibitem{rozynek} J. Rozynek, Physica A \textbf{440}, 27 (2015).

\bibitem{Biro2015} T. S Bir\'{o}, K. M Shen and B. W Zhang, Physica A \textbf{428}, 410 (2015).

\bibitem{Burrows:1986me}
  A.~Burrows and J.~M.~Lattimer,  Astrophys.\ J.\  {\bf 307}, 178
  (1986).

\bibitem{urca} J. M. Lattimer, C. J. Pethick, M. Prakash, and P. Haensel, Phys.
Rev. Lett. 66, 2701 (1991).

\bibitem{klaen06} T. Kl\"{a}hn, {\it et al.}, Phys. Rev. C 74, 035802 (2006).

\bibitem{urca2} M. Prakash,  J. M. Lattimer, and C. J. Pethick,
Astrophys. J. 390, L77 (1992).

\bibitem{bps} G. Baym, C. Pethick, and D. Sutherland, Astrophys. J. 170, 299
(1971).

\bibitem{tov} R. C. Tolman,  Phys.\ Rev.\  {\bf 55}, 364 (1939);
J. R. Oppenheimer and G. M. Volkoff, Phys.\ Rev.\  {\bf 55}, 374 (1939).

\bibitem{Hebeler} K. Hebeler, J. M. Lattimer, C. J. Pethick and A. Schwenk, Phys. Rev. Lett. \textbf{105} (2010) 161102.

\bibitem{guillot} S. Guillot, M. Servillat, N. A. Webb and R. E. Rutledge, ApJ \textbf{772} (2013) 7.

\bibitem{Lattimer2013} J. M. Lattimer and A. W. Steiner, Astrophys. J. \textbf{784}, 123 (2014).

\bibitem{rafael11} Rafael Cavagnoli, Constan\c ca Provid\^encia, and Debora P. Menezes, Phys. Rev. C 83, 045201 (2011).

\end{thebibliography}
\end{document}